\documentclass[letter,twocolumn]{jpsj3} 
\title{
Long-Range Order of the Three-Sublattice Structure 
in the $S=1$ Heisenberg Antiferromagnet 
on a Spatially Anisotropic Triangular Lattice}
\catcode`\@=11
\def\simle{\mathrel{\mathpalette\@versim<}}   
\def\simge{\mathrel{\mathpalette\@versim>}}   
\def\@versim#1#2{\lower2.5pt\vbox{\baselineskip0pt \lineskip-.5pt
   \ialign{$\m@th#1\hfil##\hfil$\crcr#2\crcr\sim\crcr}}}
\catcode`\@=12

\author{Hiroki \textsc{Nakano}$^{1}$
\thanks{E-mail address: hnakano@sci.u-hyogo.ac.jp}, 
Synge \textsc{Todo}$^{2}$
\thanks{E-mail address: wistaria@issp.u-tokyo.ac.jp}, 
and 
T\^oru \textsc{Sakai}$^{1,3}$
\thanks{E-mail address: sakai@spring8.or.jp}
}

\inst{$^{1}$Graduate School of Material Science, University of Hyogo,
Kouto 3-2-1, Kamigori, Ako-gun, Hyogo 678-1297, Japan \\
$^{2}$
Institute for Solid State Physics, University of Tokyo, 
7-1-26-R501 Port Island South, Kobe 650-0047, Japan \\
$^{3}$
Japan Atomic Energy Agency, SPring-8, 
Kouto 1-1-1, Sayo, Hyogo 679-5148, Japan
}

\recdate{\today}

\abst{We study 
the $S=1$ Heisenberg antiferromagnet 
on a spatially anisotropic triangular lattice 
by the numerical diagonalization method. 
We examine the stability of the long-range order 
of a three-sublattice structure observed 
in the isotropic system between the isotropic case 
and the case of isolated one-dimensional chains. 
It is found that the long-range-ordered ground state 
with this structure 
exists in the range of $0.7 \simle J_{2}/J_{1} \le 1$, 
where $J_{1}$ is the interaction amplitude along 
the chains and $J_{2}$ is the amplitude of other interactions. 
}

\kword{Antiferromagnetic Heisenberg spin model,
triangular lattice, frustration, spatial anisotropy, 
numerical diagonalization method, Lanczos method}

\begin{document}
\maketitle

Frustrated magnets have attracted much attention 
from the viewpoint of realizing exotic quantum states 
and phase transitions that occur between such states. 
One such magnet is 
the triangular-lattice Heisenberg antiferromagnet. 
Much effort has been devoted to studies of the $S=1/2$ model, 
particularly since Anderson\cite{Anderson} pointed out 
that this model is a possible candidate 
for the realization of the spin liquid ground state 
due to magnetic frustrations. 
Now, it is widely believed that 
the ground state of the $S=1/2$ model 
has a long-range order (LRO) with a small three-sublattice 
magnetization\cite{Huse_Elser,Jolicour_Guillou,Singh_Huse,
BBernu_CLhuillier_LPierre_PRL,BBernu_CLhuillier_LPierre_PRB}. 
This spin structure is also called the 120 deg structure 
from the directions of neighboring spins. 
However, estimating the magnetic order 
of a three-sublattice structure (3SS) 
quantitatively by direct methods  
that are unbiased beyond any approximation or variational method 
is still difficult even now.  
Numerical diagonalization data of small finite-size (FS) clusters 
of up to 36 sites of the $S=1/2$ 
model\cite{BBernu_CLhuillier_LPierre_PRL
,BBernu_CLhuillier_LPierre_PRB,PWLeung_KJRunge} were examined
\cite{comm_tri_39}.  
However, Leung and Runge\cite{PWLeung_KJRunge} and 
Bernu {\it et al.}\cite{BBernu_CLhuillier_LPierre_PRB} reported 
the subtlety in quantitative extrapolation 
from their diagonalization data of spin correlation functions 
for observing the order directly. 

\begin{figure}[htb]
\begin{center}
\includegraphics[width=7cm]{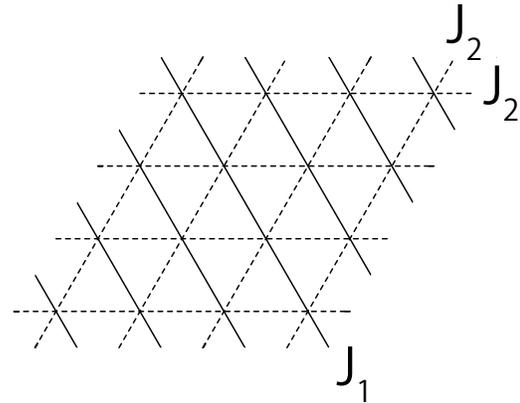}
\end{center}
\caption{Spatial anisotropic triangular lattice.  
}
\label{fig1}
\end{figure}

Since the LRO occurs 
as a consequence of the delicate balance 
of the frustrated situation,  
the instability of this order is an attractive issue 
in the case when the model includes factors 
additional to the triangular-lattice Heisenberg antiferromagnet. 
One such factor is 
the spatial anisotropy of interactions in the system. 
Let us consider that the interaction 
along a specific direction ($J_1$) 
is different from the interaction 
along the other two directions ($J_2$) 
among three equivalent directions 
in the isotropic case (see Fig.~\ref{fig1}); 
here, we focus our attention on the range 
of $0 \le J_2/J_1 \le 1$. 
In the case of $J_2=0$, the system is reduced 
to isolated spin chains 
with an antiferromagnetic interaction, in which 
the LRO does not occur 
owing to the one-dimensionality. 
Therefore, the order must disappear at a point in this range 
of $0 \le J_2/J_1 \le 1$. 

In the classical case of $S=\infty$, 
the spiral ground state is realized 
in the range of $0 \le J_2/J_1 \le 1$; 
the behavior is characterized by the canting angle 
between neighboring spins. 
In the quantum case, however, 
not only the canting angle but also 
sublattice magnetization in the spiral state 
can be changed due to quantum fluctuation; 
the situation is different from the classical case. 
The $S=1/2$ case has also been 
tackled\cite{WZheng_RHMcKenzie_RPSingh,
SYunoki_SSollera,OAStarykh_LBalents,DHeidarian_SSollera,
JReuther_RThomale,
AWeichselbaum_SRWhite,SGhamari_CKallin_SSLee_ESSorensen,KHarada}. 
On one hand, there are various reports 
concerning the range of the spiral state depending 
on the method\cite{comment_spiral_phase}. 
On the other hand, it was shown 
by the functional renormalization group approach 
that the LRO in the isotropic case disappears quickly 
when the anisotropy is switched on\cite{JReuther_RThomale}. 
Our understanding of this $S=1/2$ model has not reached 
a consensus so far. 

In this letter, we examine 
the $S=1$ triangular-lattice Heisenberg antiferromagnet 
with a spatial anisotropy. 
One major difference of the $S=1$ case from the $S=1/2$ case 
is that the ground state at $J_{2}=0$ in the $S=1$ case shows 
the Haldane gap, whereas that in the $S=1/2$ case is gapless 
in the spin excitation. 
It is a nontrivial issue how the existence of the gap 
affects the magnetically ordered phase, 
which is supposed to be realized in the isotropic case.  
Only two approximations\cite{Pardini_Singh,Li_Bishop} 
have been attempted to elucidate the $S=1$ case. 
By a linked-cluster series expansion method\cite{Pardini_Singh}, 
it was shown that 
the ratio $J_{2}/J_{1}$ at the boundary 
between the magnetically ordered phase of the spiral state 
and the disordered phase is 0.33. 
In ref. \ref{Li_Bishop}, it was shown by the coupled cluster method 
that the ordered phase spreads over 
a range of as low as $J_{2}/J_{1}=0.25$. 
In order to obtain an understanding 
of what beyond any approximation, 
direct numerical simulations have become 
increasingly important. 
However, the quantum Monte Carlo method cannot be applied 
to the analysis of the triangular-lattice antiferromagnet 
owing to the so-called negative-sign problem 
from the frustration in the system. 
Although the density matrix renormalization method 
is powerful for the analysis of one-dimensional systems 
irrespective of whether or not frustration exists, 
this method is much less effective for the analysis of 
two-dimensional systems such 
as a triangular-lattice antiferromagnet.  
Under these circumstances, 
the numerically exact diagonalization method 
based on the Lanczos algorithm is applicable 
to the triangular-lattice antiferromagnet 
irrespective of the spatial dimensionality and 
the existence of magnetic frustration. 
The purpose of the present study 
is to extract information on the magnetic LRO 
from spin correlation functions calculated 
by the Lanczos diagonalization method 
in the $S=1$ FS clusters.  
The first aim is that the LRO 
of the three-sublattice structure is quantitatively 
confirmed in the isotropic case of $J_{2}=J_{1}$ 
by the direct simulation of Lanzcos diagonalization 
in contrast with the situation mentioned above, 
for which no such confirmation in the $S=1/2$ case 
has been successfully made yet. 
The further progress in this study 
clarifies how the LRO 
behaves in the anisotropic case 
without having a biased guess based on the argument 
of the classical system. 

Here, we study the Hamiltonian given by 
\begin{equation}
{\cal H}=J_1 \sum_{\langle i,j \rangle} \mbox{\boldmath $S$}_{i}\cdot \mbox{\boldmath $S$}_{j}
+ J_2 \sum_{[ i,j ]} \mbox{\boldmath $S$}_{i}\cdot \mbox{\boldmath $S$}_{j} , 
\label{H_aniso_tri}
\end{equation}
where $\mbox{\boldmath $S$}_{i}$ 
denotes the $S=1$ spin operator. 
The sum runs over all nearest-neighbor pairs 
having the antiferromagnetic interaction of $J_{1}$ or $J_{2}$. 
Energies are measured in units of $J_{1}$; 
hereafter, we set $J_{1}=1$.  
The number of spin sites is denoted by $N_{\rm s}$. 
We take $N_{\rm s}/3$ to be an integer. 
\begin{figure}[htb]
\begin{center}
\includegraphics[width=7cm]{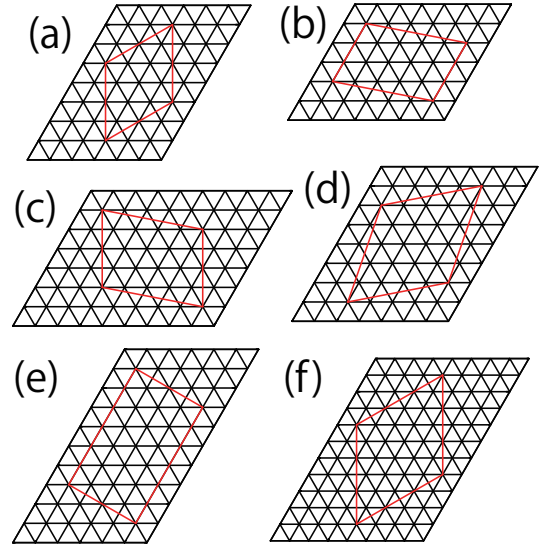}
\end{center}
\caption{Shapes of FS clusters 
of triangular lattice for 
(a) $N_{\rm s}=12$, 
(b) $N_{\rm s}=15$, 
(c) $N_{\rm s}=18$, 
(d) $N_{\rm s}=21$, 
(e) $N_{\rm s}=24$, and 
(f) $N_{\rm s}=27$.  
}
\label{fig2}
\end{figure}
In all cases of $N_{\rm s}=$12, 21, and 27, we have 
a rhombic cluster having an interior angle of $\pi/3$, 
at which the two-dimensionality may be well captured 
within FS clusters, 
although only nonrhombic clusters can be formed 
in the cases of $N_{\rm s}=$15, 18, and 24 (see Fig.~\ref{fig2}). 
Note that we exclude the cases 
in which a system of three-site chains is formed 
only by $J_{1}$ bonds 
at the point of $J_{2}=0$. 
We impose the periodic boundary condition 
for FS clusters. 
We calculate the lowest energy of ${\cal H}$ 
in the subspace divided by $\sum_{j} S_{j}^z=M$. 
The energy is denoted by $E(N_{\rm s},M)$. 
We also calculate 
the correlation function $\langle S_{i}^{z} S_{j}^{z} \rangle$ 
to observe the LRO.  

The weak point of the Lanczos diagonalization method 
is that only small clusters can be treated 
owing to the exponential increase 
in the dimension of the Hamiltonian matrix 
with respect to the system size. 
To overcome this problem, 
we have carried out parallel calculations\cite{comm_performance} 
using the MPI-parallelized code\cite{comments2para} 
to treat system sizes as large as possible. 
The case of $N_{\rm s}=$27 is the largest size 
in this study\cite{comm_max_size}. 
In this case, the largest number of dimensions is 
712~070~156~203 for the subspace of $M=0$. 
Note that this number of dimensions is the largest 
among those reported in numerical diagonalization studies 
of quantum lattice systems, 
to the best of our knowledge\cite{comment_dim}. 

\begin{table*}[hbt]
\caption{List of numerical data of the ground-state energy 
$e_{g}=E(N_{\rm s},M=0          )/N_{\rm s} $, 
the singlet-triplet energy difference
$\Delta_{N_{\rm s}}=E(N_{\rm s},M=1)-E(N_{\rm s},M=0)$, 
and correlation functions 
for the rhombic clusters. 
The distance between sites $i$ and $j$ is denoted by $|r_{ij}|$. 
}
\label{t1}
\begin{center}
\begin{tabular}{rllrrrrr}
\hline
\multicolumn{1}{c}{$N_{\rm s}$} & 
\multicolumn{1}{c}{$-e_{\rm g}/J_{1}$} & 
\multicolumn{1}{c}{$\Delta_{N_{\rm s}}/J_{1}$} 
& $\langle S_{i}^{z} S_{j}^{z} \rangle $ & & & & 
\\
 &  &  
& $|r_{ij}|=1$ & $|r_{ij}|=\frac{\sqrt{3}}{2}$ & $|r_{ij}|=2$ & $|r_{ij}|=\sqrt{7} $ & $|r_{ij}|=3$  
\\
\hline
27 &  1.870821364074 & 0.348963468544 
&-0.207869 & 0.278656 &-0.125503 &-0.140997 &0.253805  \\
21 &  1.882070035640 & 0.441247764571 
&-0.209119 & 0.286580 &-0.138420 &-0.150456 &   \\
12 &  1.959106953374 & 0.751321679528 
& -0.217679 & 0.303220 &-0.135128 & &   \\
\hline
\end{tabular}
\end{center}
\end{table*}

Now, we present our results 
of the $S=1$ triangular-lattice antiferromagnet 
for the FS rhombic clusters including 
the case of $N_{\rm s}=27$; 
these results are listed in Table~\ref{t1}. 
\begin{figure}[htb]
\begin{center}
\includegraphics[width=7cm]{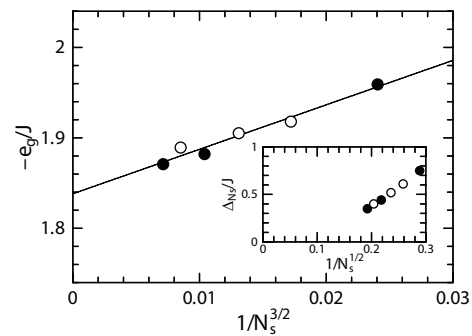}
\end{center}
\caption{
Ground-state energy per site as a function of $N_{\rm s}^{3/2}$. 
In the inset, the $N_{\rm s}^{1/2}$ dependence 
of the spin excitation gap is shown. 
Closed (open) circles represent 
data for rhombic (nonrhombic) clusters. 
}
\label{fig3}
\end{figure}
In Fig.~\ref{fig3}, the ground-state energy per site 
$e_{g}=E(N_{\rm s},M=0          )/N_{\rm s} $ 
is plotted as a function of $N_{\rm s}^{3/2}$. 
A linear least-squares fitting gives 
$-e_{\rm g}/J_{1}=1.838 \pm 0.007$, 
which is in agreement with the results 
of the spin-wave approximation\cite{Jolicour_Guillou} and 
coupled cluster method\cite{Li_Bishop}. 
The inset of Fig.~\ref{fig3} depicts 
the singlet-triplet energy difference
$\Delta_{N_{\rm s}}=E(N_{\rm s},M=1)-E(N_{\rm s},M=0)$. 
The excitation gap seems to vanish in the thermodynamic limit,
which is consistent with the existence 
of the LRO.  
These results suggest that the FS clusters shown 
in Fig.~\ref{fig2} are an appropriate series 
for analyzing our data of correlation functions. 

Next, we examine the magnetic order 
of the triangular-lattice antiferromagnet 
from FS data of 
not only rhombic but also nonrhombic clusters. 
In the isotropic case, let us focus our attention on 
the signs of $\langle S_{i}^{z} S_{j}^{z} \rangle $ in Table~\ref{t1}. 
It is possible to group all the spin sites by 
whether $\langle S_{i}^{z} S_{j}^{z} \rangle $ is 
positive or negative 
so that sites $i$ and $j$ belong to a common group 
for a positive $\langle S_{i}^{z} S_{j}^{z} \rangle $. 
One finds that the grouping divides all the spin sites 
into three equivalent sublattices. 
Thus, we evaluate the quantity defined as 
\begin{equation}
m^{\rm sq}_{\rm diag}=
\frac{1}{N_{\alpha}} \sum_{\alpha}
\frac{1}{N_{\rm s}} 
\sum_{i}
\frac{1}{N_{\rm s}/3-1} 
{\sum_{j\in A_{i}}}^{\prime} 
\langle \mbox{\boldmath $S$}_{i} \cdot 
\mbox{\boldmath $S$}_{j} \rangle , 
\label{m4order}
\end{equation}
where $\langle \mbox{\boldmath $S$}_{i} \cdot 
\mbox{\boldmath $S$}_{j} \rangle$ is evaluated by 
$3 \langle S_{i}^{z} S_{j}^{z} \rangle $, 
some of which are obtained from the values 
of spin correlation functions presented in Table~\ref{t1}, 
by taking into account the isotropic interactions 
in the spin space.
Here, the prime at the sum 
means that $j$ runs over 
the sublattice $A_{i}$ including $i$ 
while excluding the case of $j=i$. 
Here, $\alpha$ denotes the label of directions 
concerned which is chosen as $J_{1}$ among the three directions; 
we take the average with respect to the direction 
in order to take the nonrhombic cases into account. 
The quantity $m^{\rm sq}_{\rm diag}$ corresponds 
to the squared sublattice magnetization 
in the thermodynamic limit 
when the system is magnetically ordered. 
It is known that, in the isotropic case, 
the spin-wave approximation\cite{Jolicour_Guillou} 
estimates the sublattice magnetization to be 
$m_{\rm sw} = S - \Delta_{0} + O\left[ 1/S \right]$ 
with $\Delta_{0}= 0.261$. 
For $S=1$, $m_{\rm sw}^2 = 0.546$, 
which will be compared with the value extrapolated 
from our FS $m^{\rm sq}_{\rm diag}$ 
in the isotropic case. 
Note that the quantity $m^{\rm sq}_{\rm diag}$ is the same 
as eq.~({2.11d}) in ref.~\ref{PWLeung_KJRunge}. 
Note also that the analysis of $m^{\rm sq}_{\rm diag}$ does not 
directly provide us with information on the relationship 
between spins in different sublattices. 

\begin{figure}[htb]
\begin{center}
\includegraphics[width=7cm]{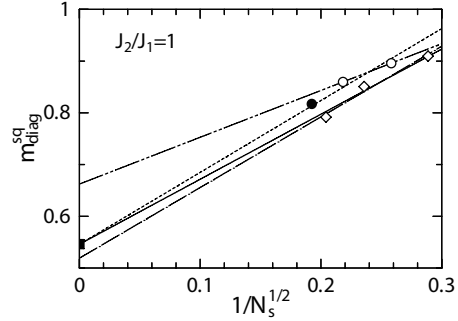}
\end{center}
\caption{
Extrapolation analysis of FS results for $m^{\rm sq}_{\rm diag}$ 
in the isotropic case. 
Open diamonds denote 
the results for $N_{\rm s}=12$, 18, and 24. 
Open circles represent 
the results for $N_{\rm s}=15$ and 21. 
The closed circle represents that for $N_{\rm s}=27$, 
which is the largest system in this study. 
The result of the spin-wave approximation 
is plotted by the closed square at the ordinate. 
See text regarding linear lines. 
}
\label{fig4}
\end{figure}

We depict $m^{\rm sq}_{\rm diag}$ in Fig.~{\ref{fig4}} 
as a function of $1/N_{\rm s}^{1/2}$. 
One can see oscillating behavior 
in the system size dependence 
with respect to whether $N_{\rm s}$ is odd or even; 
thus, the data series of odd $N_{\rm s}$ and even $N_{\rm s}$ 
should be treated separately from each other. 
Note here that 
each series of $m^{\rm sq}_{\rm diag}$ should converge to a common value 
in the thermodynamic limit. 
Then, we perform a linear least-squares fitting to each series 
under the constraint condition of a common intercept.  
From the analysis without $N_{\rm s}=27$, 
we obtain the fitting lines 
in Fig.~\ref{fig4}, 
drawn as the solid line for the data of $N_{\rm s}=12$, 18, and 24 
and as the dotted line for the data of $N_{\rm s}=15$ and 21.  
The common intercept is obtained to be 
\begin{equation}
0.546 \pm 0.051,   
\label{analysis_without_27}
\end{equation}
as an extrapolated value. 
From the analysis with $N_{\rm s}=27$, 
we have the common intercept 
\begin{equation}
0.551 \pm 0.035, 
\label{analysis_with_27}
\end{equation}
which is in agreement 
with the result~(\ref{analysis_without_27}). 
Note that both results~(\ref{analysis_without_27}) and 
(\ref{analysis_with_27}) are in agreement 
with the spin-wave result mentioned above. 
Therefore, the constraint-condition analysis 
can provide us with a reliable value from extrapolation 
even without $N_{\rm s}=27$. 
In Fig.~\ref{fig4}, the single-dotted and 
double-dotted chain lines are also drawn 
for even $N_{\rm s}$ and odd $N_{\rm s}$, respectively, 
as simple linear least-squares fittings 
without considering the constraint condition. 
One finds that the intercept 
for even $N_{\rm s}$ underestimates 
and that for odd $N_{\rm s}$ overestimates 
the squared sublattice magnetization 
with which the result 
of the constraint-condition analysis is in agreement. 

\begin{figure}[htb]
\begin{center}
\includegraphics[width=7cm]{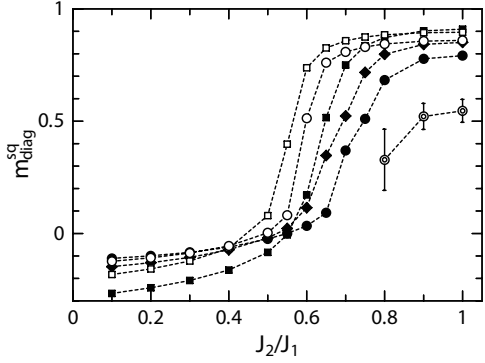}
\end{center}
\caption{
FS results for $m^{\rm sq}_{\rm diag}$. 
Closed squares, diamonds, and circles denote 
the results for $N_{\rm s}=12$, 18, and 24, respectively. 
Open squares and circles represent 
the results for $N_{\rm s}=15$ and 21, respectively. 
Double circles with the error bar 
denote extrapolated results 
from the constraint-condition analysis 
of data up to $N_{\rm s}=24$ for $J_{2}/J_{1}=0.8$, 0.9, and 1. 
}
\label{fig5}
\end{figure}

\begin{figure}[htb]
\begin{center}
\includegraphics[width=7cm]{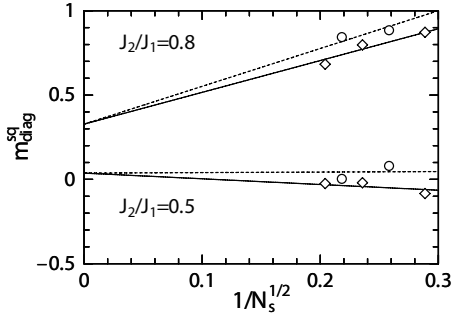}
\end{center}
\caption{
Extrapolation analysis of FS results for $m^{\rm sq}_{\rm diag}$ 
in the anisotropic cases of $J_2/J_1=0.5$ and 0.8. 
The procedure of the extrapolation is 
the same as that in Fig.~\ref{fig4}. 
}
\label{fig6}
\end{figure}

Next, we examine the anisotropic case of $J_{2}/J_{1} < 1$; 
the FS result of $m^{\rm sq}_{\rm diag}$ 
for each $N_{\rm s}$ 
as a function of $J_{2}/J_{1}$ is shown in Fig.~{\ref{fig5}}. 
One can observe that the magnetic order decreases rapidly 
as $J_2/J_1$ is decreased.  
We perform the constraint-condition analysis 
in the cases of $J_2/J_1 = 0.5$ and 0.8; 
the analysis for extrapolation is shown in Fig.~\ref{fig6}. 
The common intercept was obtained 
to be $0.04 \pm 0.17$ for $J_2/J_1 = 0.5$, 
which indicates that the LRO of the 3SS disappears. 
On the other hand, we obtain a positive common intercept 
for $J_2/J_1 = 0.8$; 
the result of $J_2/J_1 = 0.8$ is shown in Fig.~\ref{fig5} 
together with the result in the case of $J_2/J_1 = 0.9$ and 
the result (\ref{analysis_without_27}) of $J_2/J_1 = 1$.  
These results suggest that 
the LRO of the 3SS 
survives in a range having a nonzero width. 

\begin{figure}[htb]
\begin{center}
\includegraphics[width=7cm]{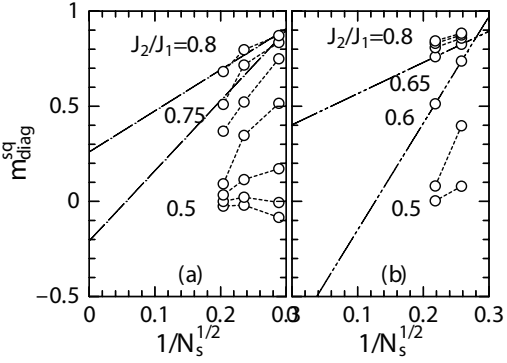}
\end{center}
\caption{
Extrapolation analysis for $m^{\rm sq}_{\rm diag}$ 
without the constraint condition
in the anisotropic cases between $J_2/J_1=0.8$ and 0.5 
with 0.05 intervals for (a) even-$N_{\rm s}$ results 
and (b) odd-$N_{\rm s}$ results.
}
\label{fig7}
\end{figure}

Between $J_2/J_1 = 0.5$ and 0.8, 
unfortunately, the constraint-condition analysis gives 
a quite large error or a negative common intercept 
owing to a severe FS effect; 
it is not so easy to estimate 
the critial value of $J_{2}/J_{1}$  
between the region where the LRO 
of the 3SS survives and 
the region where it disappears. 
In the analysis of only even $N_{\rm s}$ 
shown in Fig.~\ref{fig7}(a), 
the intercept at the ordinate changes its sign 
between $J_2/J_1 = 0.75$ and 0.8. 
From the observation of the isotropic case 
in Fig.~\ref{fig4}, 
$J_2/J_1$ at the vanishing intercept is 
considered to be overestimated 
as the critial value of $J_{2}/J_{1}$ 
from the result in Fig.~\ref{fig7}(a). 
On the other hand, the analysis 
of odd $N_{\rm s}$ shown in Fig.~\ref{fig7}(b) indicates 
that the intercept at the ordinate 
changes its sign between $J_2/J_1 = 0.6$ and 0.65; 
$J_2/J_1$ at the vanishing intercept is 
considered to be underestimated 
as the critial value of $J_{2}/J_{1}$ 
from the result in Fig.~\ref{fig7}(b). 
Therefore, it is reasonable to consider that 
$J_{2}/J_{1}$ at the boundary is $0.7 \pm 0.1$.

In summary, we studied the ground state of 
an $S=1$ Heisenberg antiferromagnet on a triangular lattice 
with spatial anisotropy 
by Lanczos diagonalization calculations 
with large-scale parallelization; 
the stability of the long-range magnetic order 
of the 3SS is examined. 
We found that in the isotropic case, 
our Lanczos diagonalization data of spin correlation functions 
are successfully extrapolated to a long-range-ordered value 
which is quantitatively consistent 
with the spin-wave approximation. 
In the present study, we concluded that, 
in the range of $0.7 \simle J_{2}/J_{1} \le 1$, 
the sublattice magnetization gradually shrinks 
as $J_{2}$ is decreased, 
while the ground state maintains the 3SS.  
The width of this region is narrower than 
those of the magnetically ordered state obtained in studies 
by approximation\cite{Pardini_Singh,Li_Bishop}. 
In future studies concerning this model, 
the magnetic structure factor as a function of wave number 
should be examined in the range near $J_{2}/J_{1} = 1$, 
which will contribute much to 
our understanding of the relationship 
between the spiral states and 
the states with the LRO of the 3SS obtained in the present study. 
In the region near $J_{2} = 0$ at which the Haldane gap exists, 
on the other hand, 
the behavior of the spin excitation gap should also be tackled. 
Since clusters up to $N_{\rm s}=24$ take 
various tilting angles, 
it is difficult to observe systematic behavior 
of the $N_{\rm s}$ dependence; 
calculations of the $N_{\rm s}=27$ system with anisotropy 
are required. 
NiGaS\cite{NiGaS} and Ba$_{3}$NiSb$_{2}$O$_{9}$\cite{HTanaka_TOno}  
are considered to be good candidate materials 
for the $S=1$ Heisenberg antiferromagnet 
on the isotropic triangular lattice, 
although experimental observations and theoretical predictions 
are not necessarily in agreement with each other in every aspect.  
Candidate $S=1$ materials in anisotropic cases 
would also contribute much to our understanding 
of magnetic phenomena due to frustration.  


\section*{Acknowledgments}
We wish to thank 
Professor S.~Miyashita, Professor Y.~Hasegawa, 
 and Dr.~T.~Okubo 
for fruitful discussions. 
This work was partly supported by Grants-in-Aid 
(Nos. 23340109, 23540388, and 24540348) 
from the Ministry of Education, Culture, Sports, Science 
and Technology of Japan. 
Some of the computations were performed using 
facilities of 
the Department of Simulation Science, 
National Institute for Fusion Science; 
Research Institute for Information Technology, 
Kyushu University; 
Center for Computational Materials Science, 
Institute for Materials Research, Tohoku University; 
The Supercomputer Center, 
Institute for Solid State Physics, The University of Tokyo;  
and Supercomputing Division, 
Information Technology Center, The University of Tokyo. 
Our computations in the largest case were carried out 
with the computational resource of Fujitsu FX10 awarded 
by the ``Large-scale HPC Challenge" Project, 
Information Technology Center, The University of Tokyo.
This work was partly supported 
by the Strategic Programs for Innovative Research, MEXT, and 
the Computational Materials Science Initiative, Japan. 
The authors would like to express their sincere thanks 
to the crew of Center for Computational Materials Science 
of the Institute for Materials Research, 
Tohoku University for their continuous support 
of the SR16000 supercomputing facilities.



\begin{thebibliography}{99}
\bibitem{comm_tri_39}
Recently, 
a 39-site system 
of an $S=1/2$ triangular-lattice antiferromagnet 
was treated in Lanczos diagonalization studies 
in refs.~\ref{JRichter_JSchulenburg_AHonecker} 
and \ref{Sakai_HN_PRBR}.
\bibitem{comment_spiral_phase}
The reported results are 
in the range down to $J_{2}/J_{1}\sim 0.85$ obtained 
by a variational Monte Carlo method\cite{DHeidarian_SSollera}, 
the whole range down to $J_{2}/J_{1}\sim 0$ obtained 
by a series expansion method\cite{WZheng_RHMcKenzie_RPSingh}, 
and 
the range down to $J_{2}/J_{1}\sim 0.7$ obtained 
using a multiscale entanglement renormalization 
ansatz\cite{KHarada}.
\bibitem{comments2para}
This parallel code was developed originally 
in the studies described 
in refs.~\ref{HN_Terai} and \ref{kgm_gap}.
\bibitem{comm_max_size}
In the $S=1$ case, the maximum size treated 
in the Lanczos diagonalization method in the previous studies 
was $N_{\rm s}=24$ in ref.~\ref{HN_Terai} 
to the best of our knowledge, 
with the criterion of 
whether the largest subspace with $M=0$ was treated. 
Reference \ref{JRichter_OGotze_RZinke_DJJFarnell_HTanaka} 
reported the $N_{\rm s}=27$ case; 
but only smaller subspaces with a large $M$ were treated.
\bibitem{comment_dim}
The record before the present study was 538~257~874~440 
in the study of the $S=1/2$ kagome-lattice Heisenberg antiferromagnet 
in ref.~\ref{kgm_gap}.
\end{thebibliography}
\end{document}